\documentclass[aps,pra,twocolumn,groupedaddress,showpacs]{revtex4}
\usepackage[dvips]{graphics,epsfig}
\usepackage{pslatex}
\usepackage{latexsym}
\usepackage{dcolumn}

\begin{document}

\preprint{LBNL-51597}

\title{Feasibility of a synchrotron storage ring for neutral polar molecules}

\author{Hiroshi Nishimura}
\email[E-mail:]{H_Nishimura@lbl.gov}
\affiliation{Mail Stop 80-101, Lawrence Berkeley National Laboratory,
University of California, Berkeley CA 94720}
\author{Glen Lambertson}
\email[E-mail:]{GRLambertson@lbl.gov}
\author{Juris G. Kalnins}
\email[E-mail:]{JGKalnins@lbl.gov}
\author{Harvey Gould}
\email[E-mail:]{HAGould@lbl.gov}
\affiliation{Mail Stop 71-259, Lawrence Berkeley National Laboratory,
University of California, Berkeley CA 94720}

\date{\today}

\begin{abstract}
Using calculations and mathematical modeling,
we demonstrate the feasibility of constructing a synchrotron storage ring
for
neutral polar molecules. The lattice is a racetrack type 3.6 m in
circumference
consisting of two of 180-degree arcs, six  bunchers, and two long straight
sections. 
Each straight section contains two triplet focusing lenses 
and space for beam injection and experiments. 
The design also includes a matched injector and a linear decelerator.
Up to 60 bunches can be loaded and simultaneously stored in the ring. 
The molecules are injected at 90 m/s but the velocity of the circulating
beam 
can be decelerated to 60 m/s after injection. 

The modeling uses deuterated ammonia ($^{14}\textrm{N}^2\textrm{H}_3$)
molecules 
in a weak-field seeking state. Beam that survives 400 turns (15 s), 
has horizontal and vertical acceptances of 35 mm-mr and 70 mm-mr
respectively, and an energy acceptance of $\pm 2$\%. 
\end{abstract}

\maketitle

\section{Introduction\label{intro}}

An electric field gradient exerts a force on the 
dipole moment of a neutral polar molecule.
The force, $F_x$, in the (transverse) $x$ direction is:
\begin{equation} \label{1}
F_x = - \frac {\partial W}{\partial x} = - \frac
{dW}{dE} \frac { \partial E}{\partial x}
\end{equation}
where $W$ is the potential energy, in an electric field 
of magnitude $E = (E_x^2 +E_y^2)^{1/2}$ (Stark effect),
of the molecule.
A similar expression may be written for the force in the $y$ direction.

It is therefore possible to accelerate 
(decelerate)\cite{bethlem99, maddi99, doyle99},
deflect, focus, and store polar molecules. 
Prior to this study, a continuous torroidal ring, 
without a buncher or matched injector was 
suggested by Auerbach et al.\cite{auerbach66}, 
analyzed by Katz\cite{katz97}, and Loesch and Schell \cite{loesch00}, 
and recently constructed by Crompvoets et al.\cite{crompvoets01},
who captured single pulses (about $10^6$ molecules) 
of deuterated ammonia ($^{14}\textrm{N}^2\textrm{H}_3$)
at 90 m/s (kinetic energy $\approx$ 9.7 K) and 
observed them for six turns (0.053 s).

 A ring with much longer storage times that can capture and store 
the high peak intensity of (the decelerated output of) 
a pulsed molecular beam jet source, could increase the available flux. 
With bunching, the energy of the stored beam can be varied 
continuously and the density of the molecules varied. 
These features make a molecular synchrotron storage ring useful for 
high-resolution spectroscopy, low energy scattering experiments, 
and for evaporative cooling \cite{hess86, ketterle96}. 
With evaporative cooling, the molecules can reach 
ultra-low temperatures where they may form quantum condensates.

Evaporative cooling requires: high densities to
thermalize the molecules by elastic scattering,  forced evaporation of the
hottest molecules, and long storage times for many repetitions of the
cooling cycle. Bunching (and focusing)
raises the density to thermalize the molecules and beam drift 
(and/or debunching) isolates the hottest molecules for removal. 

In this paper we present the results of a study showing 
the feasibility of constructing a molecular synchrotron storage ring 
that will capture large numbers of molecules in multiple bunches,
maintain the bunches, store the molecules for longer than 15 seconds and 
vary their kinetic energy.

The paper is organized as follows:
In section \ref{design} we discuss the lattice, the choice of molecular state,
the necessity of avoiding regions of very weak electric field,
and the dynamics calculations.
In section \ref{performance} we describe the 
performance of the synchrotron storage ring,
including the dynamic aperture, operating parameters, 
the effect of gravity, bunching and deceleration, and 
collisional losses.
Finally, in section \ref{system} we evaluate the 
overall performance of the storage ring by adding 
a molecular beam source, a linear decelerator, 
and an injection system.
%
\section {Storage ring design\label{design}}
\subsection {Lattice\label{thelattice}}
\begin{figure}
\includegraphics{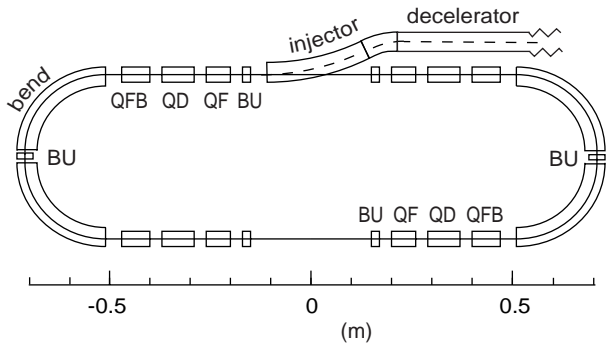}
\caption{\label{lattice} Schematic diagram of the synchrotron 
storage ring modeled in our study. 
The bend radius is 0.2 m and the circumference is 3.36 m. 
QF and QBF are horizontal focusing (vertical defocusing) lenses, QD are
horizontal defocusing (vertical focusing) lenses and BU are bunchers. 
The injector and a portion of the decelerator are also shown.
Additional details of the lattice are found in Table I, details of the
focusing lenses and bend elements are found in Table II and Fig.'s
\ref{potential}, \ref{ringfield}. 
Details of the decelerator and injector are found in 
sections \ref{decel} and \ref{inject} respectively.}
\end{figure}

\begin{figure}
\includegraphics{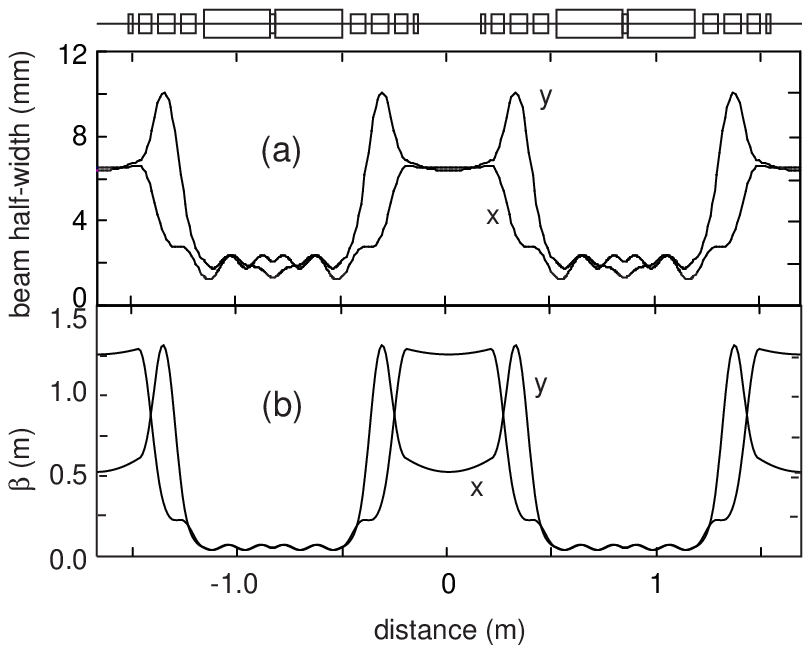}
\caption{\label{beta} Beam envelope (a) and focusing parameter 
$\beta$ (b) in the horizontal ($x$) and vertical ($y$) directions. 
$\beta$ is the distance in which the transverse (betatron) 
oscillation advances in phase by one radian.
A schematic of the lattice is shown above for location reference.
}
\end {figure}

%
Many different designs and placements of the elements 
were considered and analyzed.
Our resulting design (Fig.\ref{lattice}) is a 
racetrack configuration consisting of two of
180-degree arcs, 4 triplets of focusing lenses,
six bunchers placed in regions having low horizontal dispersion, an injector 
for loading up to 60 bunches into the ring, space in the long straight sections 
for an extractor (for example a mirror image of the injector), 
and for experiments or evaporative cooling. 
%
The lengths of the elements and drift distances between them are listed in 
Table I.
%
\begin{table}
\caption{\label{table1} Element length and placement for one-fourth of the
lattice}
\begin{ruledtabular}
     \begin{tabular}{ldc}
        Element & \mbox{Length (cm)} &\mbox{Cumulative travel (cm)} \\
        \hline
Drift   & 16.0 & 16.0 \\
Buncher (BU) &  2.0 & 18.0 \\
Drift &  2.0 & 20.0 \\
QF      &  6.0 & 26.0  \\
Drift   &  3.0 & 29.0 \\
QD      &  8.0 & 37.0 \\
Drift   &  3.0 & 40.0\\
QFB     &  7.0 & 47.0\\
Drift  &  4.0 & 51.0\\
$90^o$ Bend  & 31.4 & 82.4\\
Drift &  0.5 & 82.9\\
Half buncher (BU) &  1.0 & 83.9\\
\end{tabular}
\end{ruledtabular}
\end{table}
%
The strong combined horizontal and vertical focusing in the bend sections 
makes the ring very compact.  The triplet of straight-lens electrodes then
transforms the beam to a wider, more collimated beam that can traverse 
the 32-cm-long straight section as shown in Fig. \ref{beta}(a).
The large variation of focusing strength is apparent in 
Fig. \ref{beta}(b), the plot
of the focusing parameters $\beta_x$ and $\beta_y$ around the ring. 

\begin{figure}
\includegraphics{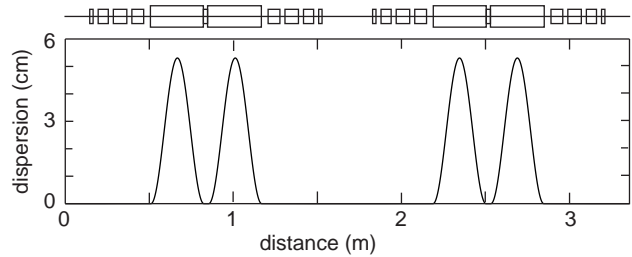}
\caption{\label{dispersion} Horizontal dispersion 
of the beam around the lattice.
Low dispersion at the bunchers prevents the 
development of synchro-betatron oscillations.}
\end {figure}

The horizontal betatron phase advance in the 180-degree bending region 
has been made equal to $4\pi$ in order to have zero dispersion at the
bunching electrode, as shown in Fig. \ref{dispersion}.  
The very low dispersion at the bunching electrodes 
prevents the development of strong synchro-betatron coupling and 
preserves the momentum acceptance. 
%
\subsection {Molecular state\label{state}}
%
\begin{figure}
\includegraphics{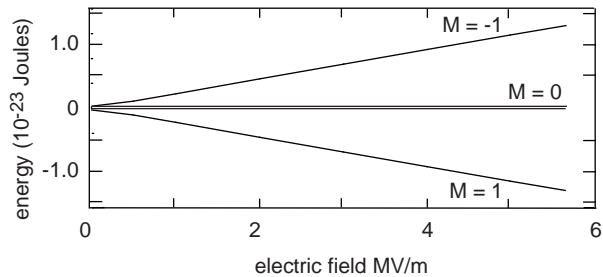}
\caption{\label{ND3} 
Stark effect in the $|J,K> = |1,1> $ levels of
$^{14}\textrm{N}^2\textrm{H}_3$.
The $M = -1$ level is weak-field seeking and used for this study. 
Inversion splitting, dipole moment and rotational
constant are taken from Townes and Schawlow \cite{townes55}.Hyperfine
splitting and the Stark effect in the $10^{-24}$ Joule inversion-split $M = 0$
levels are neglected. }.
\end{figure}
%
Many different molecules and kinetic energies 
could be used for our feasibility study.
We selected 90 m/s $^{14}\textrm{N}^2\textrm{H}_3$ 
molecules in the weak-field seeking upper-inversion level 
(lowest vibrational state) $|J,K,M> = |1,1,-1> $ (Fig. \ref{ND3})
because they had been previously used by 
Crompvoets et al.\cite{crompvoets01}, 
and because the focusing properties of weak-field seeking states 
make the design and feasability assessment easier and the ring more
compact.

Molecules in weak-field seeking states 
(molecules whose potential energy increases 
in an electric field) can be focused 
simultaneously in both transverse directions, whereas
molecules in strong-field seeking states 
can be focused in only one transverse direction 
while defocusing in the other transverse direction. 

Fringe fields seen by the molecules upon entering and exiting the bending,
focusing, and bunching elements will also exert transverse forces on the
molecules. 
A molecule in a weak-field seeking state traveling in the  
$z$ direction will experience a net focusing effect 
in the vertical ($y$) direction upon entering or 
exiting an electric field produced by horizontal ($x, z$) plane parallel
electrodes.
A molecule in a strong-field seeking state will defocus. 

Except for the rotational state $J = 0$ 
(which is always strong field seeking), 
each rotational state, $J$, contains $M$ components which 
are partially or completely degenerate in zero electric field.
The different $|M|$ states (or $M$ states as in Fig. \ref{ND3})
have different Stark shifts and hence experience different forces in an
electric field gradient.  
If the molecules in a ring repeatedly enter regions of weak and
direction-changing electric fields, transitions between the different $M$
components will take place (Majorana transitions), 
leading to a loss of the molecules\cite{reuss88, harland99, kajita01}. 
Our method of avoiding this problem is described below.
%
\subsection {Avoiding weak-field regions\label{avoiding}}
\begin{figure}
\includegraphics{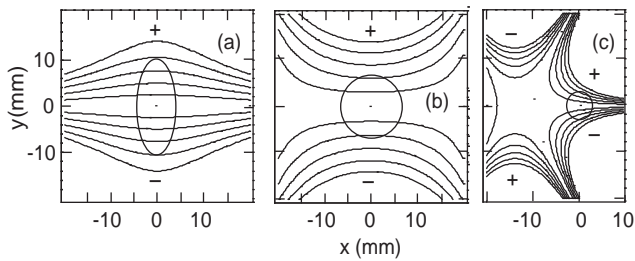}
\caption{\label{potential} 
Maps of  (truncated) equipotentials in $x$ and $y$ of 
(a) F lenses, QF and QFB (horizontal focusing, vertical defocusing), 
(b) D lenses, QD (horizontal defocusing, vertical focusing),
and (c) bend elements (horizontal and vertical focusing). 
The ellipse and circles show the approximate size 
of the beam envelope in the element. 
Note that the electric field direction at the orbit of the molecules 
is always in the vertical direction. 
The actual electrode can be fabricated to lie along any set of equipotentials 
that are larger than the beam size. 
}
\end{figure}

In the bend sections, which use electrodes with zero electric field near their
center, the centripetal force keeps the trajectory of molecules in a strong
electric
field at the inner side of that zero (Fig. \ref{potential}c). 
And, as did Crompvoets et al.\cite{crompvoets01}, 
we take advantage of the combined  horizontal and vertical focusing.
 
\begin{figure}
\includegraphics{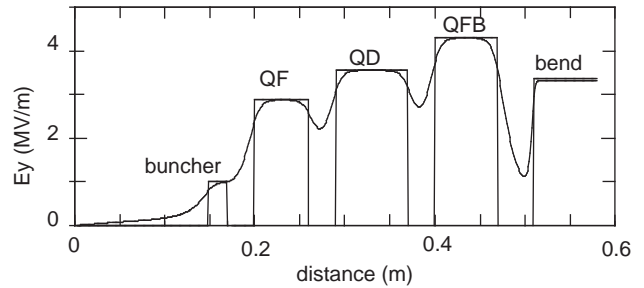}
\caption{\label{ringfield} 
Electric field magnitude around the ring. 
All electric fields are in the vertical direction and do not change sign.
The buncher field is shown for an on-energy molecule entering or exiting the
buncher (see section \ref{bunch}).
The fringe fields are part of the focusing system 
and are discussed in section \ref{fringe}.}
\end{figure}

Focusing lenses in the straight sections (Fig. \ref{potential}a,b) 
are sextupolar with a dipole field added to avoid zero field. 
The multipole coefficients of the electrodes, their gaps, 
and electric field strengths on orbit, are listed in Table II. 
With the added dipole field, these lenses focus in only one transverse
direction while defocusing in the other and 
are used in an alternating-gradient sequence \cite{kalnins02}.   
To prevent rapid changes in electric field direction, the
field direction, at the molecule's orbit throughout the ring, is vertical 
and remains unchanging in polarity (Fig. \ref{ringfield}).  
This results in having some concave and some convex lenses.

In the straight sections, away from focusing fields, 
we add a weak bias field (about one kV/m for the drift spaces) 
to maintain the quantization axis\cite{harland99}.
\begin{table}
\caption{\label{table:table2} Bending and focusing/defocusing electrodes}
 \begin{ruledtabular}
 \begin{tabular}{lccccc}
        Electrode & Length (cm) & $E_0$ (MV/m) & Half-gap (mm)&$A_2$
&$A_3$ \\
        \hline
        bend & 31.4 & 3.37 & 6 & 157 & 5667 \\
        QF & 6.0 & 2.88 & 15 &  0 & 2000 \\
        QD & 8.0 & 3.55 & 15 & 0 & -2000 \\
        QFB & 7.0 & 4.30 & 15 & 0 & 2000 \\
\end{tabular}
\end{ruledtabular}
\end{table}
%
%
\subsection {Dynamics calculation\label{calculation}}
The force on a molecule in one direction is given by Eq. \ref{1}.
The basic formulas for motions of the molecules 
are derived analytically without linearization.
Then the beam optics and dynamics are numerically calculated and
optimized by using a modified version of the program 
Goemon\cite{goemon} that was originally
developed for relativistic charged particles.
The molecules were numerically tracked through the synchrotron 
for 400 turns (15 s)  to determine the beam's dynamic stability.

The numerical integration was done in the time domain because the speed
of a molecule varies as a function of the electric field. 
The overall dimension of the lattice is chosen to 
balance  easily attainable electric fields and high transverse acceptance 
with compact (desk top) size. 
%
\subsubsection {Potential energy\label{energy}}
The potential energy, W, in an electric field, of the $|J,K> =|1,1>$ level of
 $^{14}\textrm{N}^2\textrm{H}_3$ (Fig. \ref{ND3}) is given, in units of
Joules 
(written out to avoid confusion with the quantum number "$J$") by:
 \[  W=\pm \left[\sqrt{C_1^2+C_2^{2}E^2}-C_1-C_3E^2 \right]\]
where $C_1 = 5.26\times 10^{-25} $Joules 
is half of the  $|J,K> =|1,1>$ inversion splitting (Fig. \ref{ND3}), 
$C_2 = 2.52\times 10^{-30}\textrm{Joules}\textrm{V}^{-1}\textrm{m} $, 
 $C_3 = 1.78\times10^{-38}
\textrm{JoulesV}^{-2}\textrm{m}^{2}$ and $E$ is in V/m. 
The terms involving $C_1$ and $C_2$ are taken from Ref.\cite{townes55}
to which we add a second order term to 
account for mixing of the $|J,K>$ = $|2,1>$ state
in strong electric fields\cite{amini01}.
%
\subsubsection {Electric field gradient\label{gradient}}
Within an electrode, the electric field, 
as a function of distances $x$ and $y$ 
from the reference orbit, is calculated 
from a scaler potential $\psi$ by
$E = -\nabla{\psi(x,y)}$ and $\psi$ is taken 
to be uniform in the longitudinal ($z$) direction within the
electrode length. The fringe fields, at the ends of the electrodes, where
the electric field is changing in the $z$ direction of motion 
produce additional transverse forces.  These are evaluated separately, and
are described in section \ref{fringe}. 

In a straight (focusing/defocusing) electrode, the potential is a
combination of dipole and sextupole terms \cite{kalnins02} given by  
 $\psi_s= -E_0[y+ A_3 (x^2 y-\frac{1}{3}y^3)]$
where $E_0$ is the central electric field, $A_3$ is the 
sextupole coefficient (see Table II) and the dipole coefficient has been set
equal to 1.

In a bending electrode of constant radius $\rho$, $\psi$ is a
combination of dipole, quadrupole, and sextupole terms given by:
\[\psi_b = -E_b\left[y + yB_2 \ln\left(1+\frac{x}{\rho}\right)+B_3
J_0[k(\rho+x)]\sinh(ky)\right]\]
Relating this to the Cartesian multipoles, we find:
$E_b = E_0 \left[1+ 2A_3/k^2 \right]$,
$B_2 = \rho A_2 \left[k^2/k^2 +2A_3 \right]$,
 where $A_2$ is the quadrupole coefficient,
\[B_3 = - \frac{1}{k J_0(k \rho)} \frac{2A_3}{k^2 + 2A_3},\]
$J_0$ and $J_1$ are the $0^{th}$ and $1^{st}$ order Bessel functions 
and the value of $\rho k$ is the first root of $J_1(k\rho)=0$.  
An approximation for $\psi_b$ useful for 
comparison with $\psi_s$ up to third order is
 \[\psi_b =  -E_0 \left[ y+A_2 xy +
A_3 \left[ \left( 1- \frac{A_2}{2 \rho A_3} \right)x^2y-
\frac{y^3}{3}\right] \right]\]
%
\subsubsection {Fringing fields\label{fringe}}
To evaluate the transverse vertical focusing forces in the fringing
regions of the elements, a two - dimensional numerical calculation, using the
geometries of the electrodes,  
was carried out to find the electric field as a function of z on the midplane.
The magnitude of the fringe fields are shown in Fig. \ref{ringfield}.
The focusing action was then calculated analytically from the derivatives of
this field, evaluated as equivalent thin lenses 
at the ends of the focusing, bending, and bunching electrodes, and included
in the calculation of trajectories around the ring.
\subsubsection{Hamiltonian\label{hamiltonian}}
All optics, orbit, and tracking calculations were carried out using a 
second-order symplectic integrator and the Hamiltonian:
$H = H_0 +W(x,y)$.
In straight regions,
$H_{0}  = \frac{1}{2m}(P_{x}^2+P_{y}^2+P_{z}^2)$
 where $m$ is the mass of the molecule, $P_x  = mv_x$, $P_y = mv_y$, and
$P_z = mv_z$ 
 is the momentum, with $v_x$, $v_y,$ and $v_z$ the velocities in Cartesian
coordinates. 
Conservation of energy requires that $P_z$ changes 
when the molecule passes through a 
field gradient at the ends of the electrodes.

 In a bending region, the Hamiltonian $H_0$ becomes:
\[H_{0}=\frac{1}{2m}\left[P_{x}^2+P_{y}^2+\frac{P_\theta^2}{(\rho+x)^{2}}\right]\]
where  $P_\theta$ is an angular momentum, $\rho$ is a
reference bend radius, $x$ is a radial excursion with respect to $\rho$,
and $P_z = (P_\theta /\rho +x)^2$.
%
 \section{Synchrotron performance\label{performance}}
\subsection{Dynamic aperture\label{aperture}}
\begin{figure}
\includegraphics{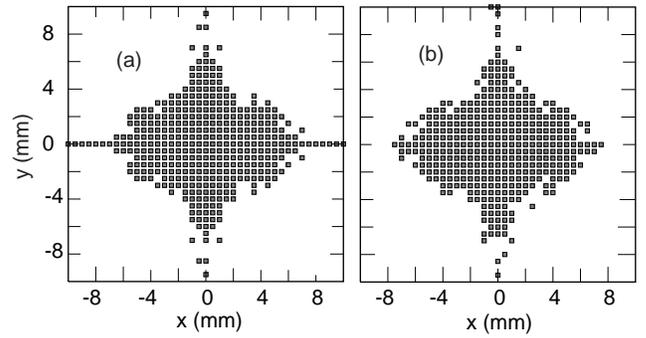}
\caption{\label{dynamic} Dynamic aperture for on-momentum molecules:
the starting coordinates for the molecules that survive 400 turns 
(a) without gravity and (b) with gravity. 
Dynamic aperture is the area, in the transverse plane, at the 
center of the  long straight section, occupied by the molecules.}
\end {figure}

 %
The survival of molecules, tracked through 400 turns, is calculated 
to determine the usable dynamic apertures, $a_x$ and $a_y$.
Fig. \ref{dynamic} shows a scatter plot of the starting coordinates 
$x$ and $y$ at center 
of the straight section of those molecules that survive the 400 turns.  
From these we calculate the acceptances, $\epsilon$, as 1/$\pi$ times the
areas  in displacement-angle phase space, $\epsilon = a^2/\beta$.
The dynamic apertures, 
values of beta at center of straight section, and acceptances are given in
Table III. 
In Table III the circulation period is 38 ms whereas 3.357/90=37.3 ms,  
reflecting the 3\% to 4\% reduction in velocity
when the molecules enter the electric fields. 
%

\begin{figure}
\includegraphics{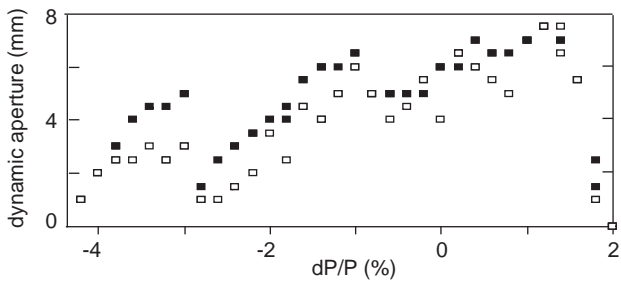}
\caption{\label{momentum} Dynamic aperture as a function of momentum
variation from the nominal tune for 
 $x$ (solid points)  and $y$ (open points)
with gravity but without bunching. The molecules in 
Fig. \ref{dynamic} appear here at dP/P = 0.}
\end {figure}

The dynamic aperture of off-momentum molecules 
determines the momentum acceptance.
Fig. \ref{momentum} shows the effect of momentum on the dynamic
aperture, 
and indicates  an acceptance of about -3\% to +1.7\%. 
We shall see later that this reduces to about $\pm2\%$ when the
bunchers are operating. 

The betatron tunes, as a function of momentum, are found 
by tracking the motions of a molecule over 512 turns.  
The chromaticities, 
are $\zeta_x = -0.0885$ and $\zeta_y = -0.0942$.
The effect of momentum-deviation upon the circulation period T, expressed
as the momentum-compaction factor $\alpha$ 
in $ \Delta T/T_0 = \alpha \Delta P/P_0$, is strong.
The  value of $\alpha$ is -0.991; this will result in prompt debunching if the
bunching voltages are turned off.

\begin{table}
\caption{\label{table:table 3} Synchrotron operating parameters}
\begin{ruledtabular}
\begin{tabular}{ll}
       \hline
           Circumference (m) & 3.357\\
           Circulation period (s) & 0.0380\\
           Velocity in free space (m/s)  & 90.0 \\
           Beta horizontal \footnotemark[1]: $\beta_x$ (m)& 1.264\\
           Beta vertical \footnotemark[1]: $\beta_y$ (m)& 0.513\\
           Horizontal dispersion\footnotemark[1] (m)& 0.001 \\
           Horizontal tune: $\nu_x$ & 5.250 \\
           Vertical tune: $\nu_y$ & 5.200 \\
           Chromaticity - horizontal: $\zeta_x$ & -0.0885 \\
           Chromaticity - vertical: $\zeta_y$ & -0.0942\\
           Momentum compaction: $\alpha$ & -0.99 \\
           Dynamic aperture - horizontal: $a_x$ (mm) & 6.5\\
           Dynamic aperture - vertical: $a_y$ (mm) & 6.0\\
           Acceptance - horizontal: $\epsilon_x$ (mm - mr) & 35\\
           Acceptance - vertical: $\epsilon_y$ (mm - mr) & 71\\
\end{tabular}
\end{ruledtabular}
\footnotetext[1]{At the center of the long straight section.}
\end{table}
%
\subsection{Effect of gravity\label{grav}}
\begin{figure}
\includegraphics{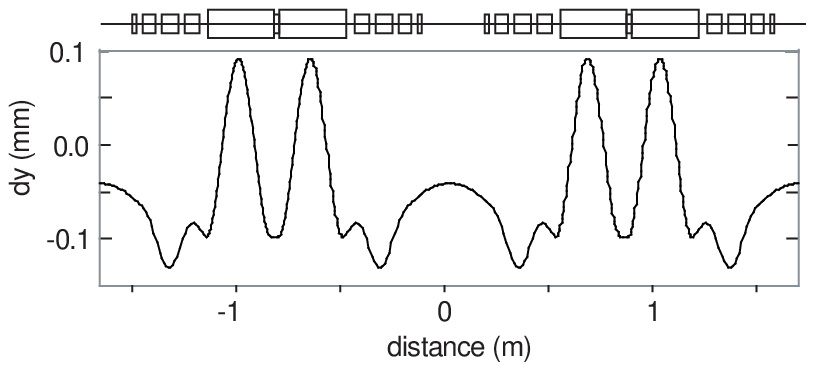}
\caption{\label{gravity} 
Closed orbit displacement (COD) from the reference orbit due to gravity. 
The swings of the COD in the bending sections are similar to the curve of
the horizontal dispersion because the vertical and horizontal 
betatron tune advances are both $4\pi$.}
\end{figure}

%
The effect of gravity is visible in the orbit because the velocity of the beam
is low.  The vertical phase advance of $4\pi$ 
(same as the horizontal phase advance), is favorable in reducing the effect 
of the gravity force on the vertical closed orbit.  
Fig. \ref{gravity} shows the vertical closed-orbit displacement
(COD) due to gravity. It ranges from -0.13 mm to 0.09 mm and is small
enough to be
neglected in the case of linear optics calculations. 
In the case of dynamic aperture calculations we have confirmed that 
gravity changes only the details of the dynamic aperture at
its edges and does not affect its core portion (Fig. \ref{dynamic}).
%
%
%
%
\begin{figure}
\includegraphics{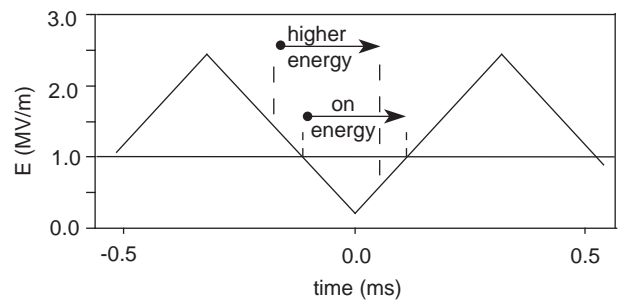}
\caption{\label{timing} 
Buncher timing for an on-energy molecule and a faster molecule.
Changing the buncher timing may be used to change the circulating
beam energy.}
\end{figure}
%
\subsection{Bunching and deceleration\label{bunch}}
%
Bunching action is provided by six sets of 
parallel-plate electrodes, each 20
mm in length with 10 mm half-gap, placed 
in low dispersion regions around the ring (Fig.\ref{lattice}, \ref{dispersion}). 
The voltage on the plates is pulsed in a  triangular waveform that produces 
a maximum electric field of 1.0 MV/m for on-energy molecules.
Each buncher is synchronized to the on-energy molecules (Fig. \ref{timing})
so that the energy lost upon entering the 
buncher is matched by the energy gained upon exit. 
A higher-energy molecule reaches the buncher at an earlier time 
when electric field gradient is higher upon entrance and
lower upon exit, resulting in a net energy reduction for a molecule
in a weak-field seeking state. 

The buncher frequency is 1.58 kHz and
allows 60 bunches around the ring, spaced 56 mm apart. 
To prevent Majorana transitions the voltage is
biased to avoid negative fields. 
The rate of change of the electric field in the bunchers can be made as large
as about 7 GVm$^{-1}$s$^{-1}$. The synchrotron tune is proportional to the
square root of the rate of change and is 0.92 at our reference value of  7
GVm$^{-1}$s$^{-1}$.

The lattice is optimized for a nominal velocity of 90 m/sec. After stacking at
this velocity, the beam can be decelerated by synchronously changing the
electrode and buncher settings. This provides an opportunity to scan the
velocity during an experiment.
The betatron tunes stay constant but the straight section becomes dispersive. 
At 63 m/sec, the horizontal transverse acceptance reduces from 35 to 13
mm-mrad and the momentum acceptance from 2.0\% to 1.2\%.
%
\begin{figure}
\includegraphics{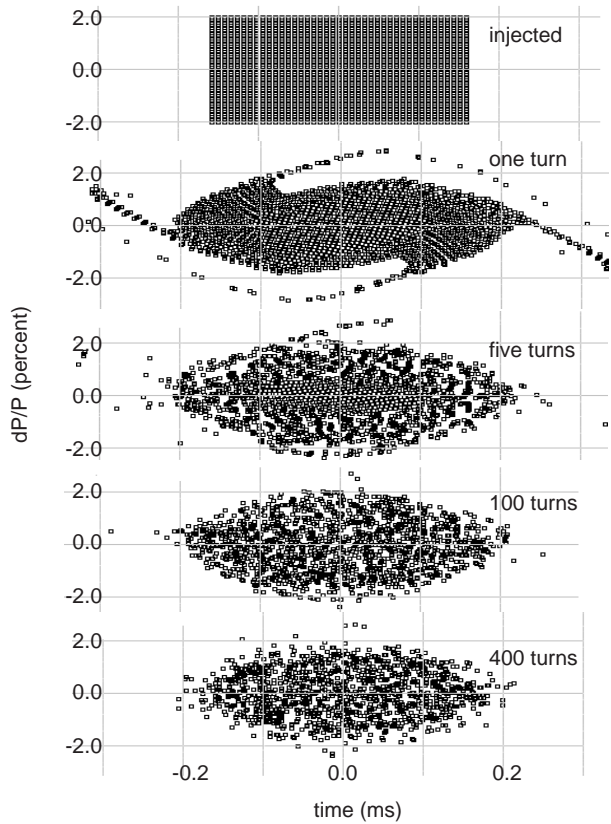}
\caption{\label{bunching} 
Scatter plot of an initially injected $\pm 158 \mu$s-long pulse 
of molecules with a $\pm2\%$ momentum spread after 1 - 400 turns. 
The velocity spread of $\pm1.8$ m/s represents an energy spread 
in the moving frame of about $\pm0.4$ K. }
\end{figure}
\subsection{Collisional losses\label{losses}}
So far, we have only examined storage ring losses associated with the limits 
of the dynamic aperture and momentum  acceptance of the lattice. 
A real storage ring will also have losses due to 
elastic and inelastic scattering of the molecules. 
Scattering by room-temperature background gasses is 
the major elastic scattering contribution and of  the gasses 
likely to be present in the ring, xenon and ammonia 
have the largest scattering coefficients. 
Assuming hard sphere binary collisions between the 
$^{14}\textrm{N}^2\textrm{H}_3$ molecules and 300 K xenon 
(leaking in from the source), 
the mean time between collisions is 15s at a (xenon) pressure 
of $1 \times 10^{-7}$ Pa ($7.5 \times 10^{-10}$ Torr), where we have used 
the equations and collision diameter in Ref.\cite{crc02}. 

For inelastic scattering between the 
$^{14}\textrm{N}^2\textrm{H}_3$ molecules, 
their relative kinetic energy spread of about 
$\pm$ 0.4 K ($\pm 5.5 \times10^{-24}$J) 
is small enough to prevent any significant 
excitation to higher rotational states. 
A molecule can however collisionally 
relax to a strong-field seeking state,
causing it to be lost from the storage ring \cite{crompvoets01}. 
Bohn \cite{bohn01} and Kajita \cite{kajita02} 
have calculated inelastic collision rates for molecules 
in a weak-field seeking state in electrostatic traps. 
They find that the loss rates can be significant, 
in some cases precluding successful evaporative cooling, 
and are influenced both by the electric fields and 
the size of the electric dipole moment. 

Thus, the use of a synchrotron storage ring 
with high densities of molecules in a weak-field 
seeking state will need to confront this issue. 
Obviously, the alternative would be to use molecules 
in the strong-field seeking rotational ground ($J = 0$) state. 
A synchrotron storage ring for molecules in 
the $J = 0$ state will be discussed in a future paper. 
%
%
\section{Synchrotron storage ring system\label{system}}
%
The overall performance of a molecular synchrotron storage ring  
also depends upon the beam delivered by the source, decelerator 
and injection line. To model performance and determine the stored 
beam from the complete system we also modeled a decelerator, 
injector and source.
%
\subsection{Linear decelerator\label{decel}}
%
\begin{figure}
\includegraphics{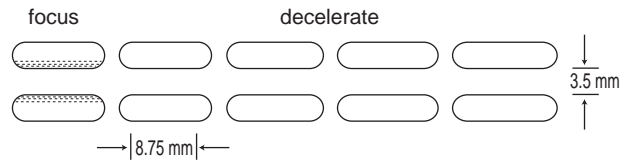}
\caption{\label{decelerator} 
Last elements of a linear decelerator showing a group of four
decelerating electrodes and one focusing lens.
The lengths of the decelerating elements and focusing lenses decrease
to match the velocity of the decelerating molecules. }
\end{figure}
%
Slow molecules, suitable for injection into a storage ring, 
can be produced by
time-varying electric field gradient deceleration, by
mechanical cancellation of the molecular beam velocity\cite{gupta99},
and possibly by buffer-gas cooling\cite{doyle95} (without magnetic
trapping).
Time-varying electric field gradient deceleration is the easiest of these
for us to model and to match to the storage ring. It has been
used by Bethlem et al. \cite{bethlem00, bethlem02} and 
Crompvoets et al. \cite{crompvoets01} to decelerate
$^{14}\textrm{N}^2\textrm{H}_3$.

Our model linear decelerator takes the 310 m/s (115 K kinetic energy) 
output of a $^{14}\textrm{N}^2\textrm{H}_3$ - seeded 
xenon pulsed-jet source (room temperature reservoir) 
and decelerates it to 90 m/s (9.75 K kinetic energy). 
The number of electrodes is set by the decrease in kinetic energy
in each electrode (equal to the change in potential energy of
the molecule entering the electric field). 
We use 79 decelerating electrodes, decreasing in length 
from 48 mm to 8.75 mm effective length to keep the 
transit time of the molecules constant through each electrode. 
We choose the overall length of the decelerator, 
3.4 m, comparable to the 3.36 m circumference of the storage ring, 
to balance high velocity acceptance with compact size.

The electric field is nearly a square wave 158 $\mu$s long 
with a repetition rate of 3.16 kHz and the maximum 
electric field is 8 MV/m in a 1.75 mm half-gap and is 
the same for all decelerating electrodes. 
After the molecules enter the electric field 
it drops to nearly zero so that the molecules that exit 
the electrode must relinquish kinetic energy to enter
the field in the next set of electrodes. 
This is done either by having successive electrodes 180 degrees 
out of phase or by using only every other bunch.
The electric field does not return completely to zero  
and the horizontal focusing elements (see below) 
have their electric field in the same direction as the 
decelerating electrodes to minimize Majorana transitions.

Bunching, as the molecules decelerate, is accomplished by having 
the electric field (in each bunching electrode) decrease, linearly over the 
158 $\mu$s from 8 MV/m to 7.76 MV/m so that the fastest molecules,
arriving early, receive the most deceleration. 
This results in a $\pm$1\% momentum spread at 90 m/s 
which falls within the momentum acceptance of the storage ring. 
 
The spacing between individual decelerating electrodes remains constant.
This fixes the fringe field which provides vertical focusing. 
The decelerating electrodes are grouped in sets of four 
as shown in Fig. \ref{decelerator}. 
After each quadruplet of decelerating electrodes is placed 
a horizontal focusing element with length appropriate 
to keep the molecules in phase with the next decelerating quadruplet. 
(One section has only three decelerating electrodes.)
The 20 focusing elements all have the same field but their 
focusing strength increases as the molecules slow. 
The lattice is, in the vertical direction, FFFFD and in the 
horizontal direction OOOOF. The overall length of the decelerator is 3.4 m. 

Upon exit from the decelerator, bunches are 7 mm long with a vertical and
horizontal half-width of 1.30 mm and 1.61 mm, respectively. 
The vertical and horizontal betas are:
$\beta_v = 56.3$ mm and $\beta_h = 86.7$ mm 
corresponding to vertical and horizontal emittances both of 30 mm-mr. 
Being less than the 71 mm-mr and 35 mm-mr vertical and 
horizontal acceptances of the synchrotron storage ring, 
this sets the overall transverse acceptance of the entire system.
%
\subsection{Injector\label{inject}}
%
\begin{figure}
\includegraphics{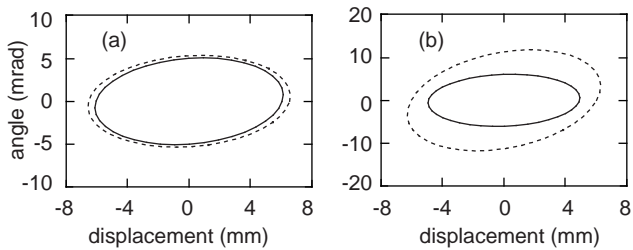}
\caption{\label{matching} 
Boundaries of injected-beam emittances (solid lines) and storage ring 
acceptances (broken lines) in the (a) horizontal and (b) vertical planes. }
\end{figure}
%
A bunched beam of molecules is injected onto 
the closed orbit of the synchrotron at the 
downstream end of a straight section.
Starting at the exit of the decelerator, this beam is guided 
along a trajectory by electric fields that focus transversely 
and transform itÕs size and divergence to match 
the vertical and horizontal acceptances of the storage ring. 

The last element must be a pulsed deflecting electrode 
that turns off after the bunch or bunches have entered.
Several different injection protocols are available:
a string of  158 $\mu$s pulses can be injected into the ring
in less than the circulation period of 38 ms, or single bunches
may be filled at any time by switching the pulsed deflecting electrode
during the interval between bunches.

The injector guide field consists of two bending sectors, 
each an arc of about $\pi/8$ rad, arranged as shown in Fig.\ref{lattice}.
In each bending sector the field is configured 
to provide equal horizontal and vertical focusing. 
The phase  advance of transverse motion is near $\pi/2$ in 
a sector; thus, it exchanges angle for displacement and changes 
the ratio of angle to displacement.
Focusing strengths are adjusted to provide the required 
match between injecting beam and ring acceptance. 
The boundaries of the beam phase space are shown in Fig.\ref{matching}. 

It was possible to choose the first sector to have radius and 
strength equal to that of the bend in the storage ring. 
The second sector must have larger aperture and radius 
and it must be pulsed to zero at the end of injection. 
Parameters of the injector guide field are shown in Table IV. 
The sum of the two inverse curves was made near zero 
so that the line of the decelerator is about parallel 
to the straight section of the synchrotron storage ring. 
\begin{table}
\caption{\label{table:table4} Injection matching fields}
\begin{ruledtabular}
\begin{tabular}{lll}
       \hline
          & \mbox{Sector 1}&\mbox{Sector 2}\\
           Electric field (MV/m) & 3.37 & 4.0 \\
          Dipole coefficient ($\textrm{m}^{-1}$) & 158.5 & 38.2\\
   Radius (m) & -0.20 & 0.69 \\
       Arc length (rad) & 0.397 & 0.426 \\
    Phase advance (rad) & 1.57 & 1.55 \\
Focus parameter, $\beta$ (m) & 0.05 & 0.19 \\
\end{tabular}
\end{ruledtabular}
\end{table}
%
\subsection{Source and intensity\label{intensity}}
To estimate the number of molecules that can be decelerated, 
injected, and stored, we assume $^{14}\textrm{N}^2\textrm{H}_3$ 
source conditions similar to those reported by 
Crompvoets et al.\cite{crompvoets01} and by 
Bethlem et al.\cite{bethlem02}: 
a pulsed jet source of 0.8\% $^{14}\textrm{N}^2\textrm{H}_3$ 
seeded in 152 kPa (1140 Torr) of xenon (reservoir temperature 300 K) 
exiting through a circular 0.80 mm dia. orifice into vacuum. 
Following Crompvoets et al. \cite{crompvoets01},
we assume that 15\% of the molecules entering 
the decelerator are in the desired state. From the formulas in 
Miller\cite{miller88}, we find a flux of $4.4 \times10^{18}$ 
$^{14}\textrm{N}^2\textrm{H}_3$ 
molecules $\textrm{sr}^{-1} \textrm{s}^{-1}$ in the desired $M$ state.

The decelerator's transverse emittance of $\pm$30 mm-mr 
and momentum spread of $\pm1$\% at 90 m/s set a 
transverse acceptance from the 310 m/s jet source beam
of $\pm9$ mm-mr and momentum spread of $\pm 0.29\%$.
If we assume an initial Gaussian velocity distribution with a 
mean of 310 m/s and a standard deviation of $\pm20\%$,  
approximately 1.1 \% of the molecules from the jet source 
fall within our decelerator momentum acceptance. With a pulse length 
of 158 $\mu$s, the intensity of the decelerated beam is 
roughly $6 \times 10^8$ molecules/pulse. 

The horizontal acceptance of the storage ring matches the emittance of the
decelerator and the vertical acceptance and momentum acceptance of the 
storage ring is about twice the emittance of the injected beam, so 
all $6 \times 10^8$ molecules/pulse should be captured and 
stored in a single bunch.  If all sixty bunches are filled (over multiple turns) 
the total stored beam is $3.6 \times 10^{10}$ molecules and the circulating
flux is
$9.5 \times 10^{11}$ molecules/s. 

\begin{acknowledgments}
We thank Swapan Chattopadhyay and 
Ying Wu for early assistance with this work. 
Work on the synchrotron storage ring is supported 
by the Director, Office of Science, of the U.S. Department of Energy, 
and work on the linear decelerator is supported by 
the Director, Office of Science,
Office of Basic Energy Sciences, of the U.S. Department of Energy;
both under Contract No. DE-AC03-76SF00098. 
\end{acknowledgments}

\bibliography{synchbib}
\end{document}